\documentclass[twocolumn,showpacs,amsmath,amssymb,prl,superscriptaddress,a4paper]{revtex4}

\usepackage{graphicx}
\usepackage{dcolumn}
\usepackage{bm}
\usepackage{anysize}
\usepackage{graphicx}
\usepackage{times,color}
\usepackage[ps2pdf]{hyperref}

\begin{document}

\preprint{APS/123-QED}

\title{Resonant excitonic emission of a single quantum dot in the Rabi regime}

\author{R. Melet}
 \email{Romain.Melet@insp.jussieu.fr}
\author{V. Voliotis}
\author{A. Enderlin}
\author{D. Roditchev}

\affiliation{Institut des NanoSciences de Paris, Universit\'{e} Pierre et Marie Curie, CNRS UMR 7588,
Campus Boucicaut, 140 Rue de Lourmel, 75015 Paris, France}

\author{X. L. Wang}
\affiliation{Nanotechnology Research Institute, National Institute of Advanced Industrial Science and Technology (AIST), Tsukuba Central 2, Tsukuba 305-8568, Japan}

\author{T. Guillet}
\affiliation{Groupe d'Etude des Semiconducteurs, Universit\'{e} Montpelier II, Case Courier 074, 34095 Montpellier Cedex 05, France}
\author{R. Grousson}
\affiliation{Institut des NanoSciences de Paris, Universit\'{e} Pierre et Marie Curie, CNRS UMR 7588,
Campus Boucicaut, 140 Rue de Lourmel, 75015 Paris, France}

\date{\today}

\begin{abstract}
We report on coherent resonant emission of the fundamental exciton state in a single semiconductor GaAs quantum dot. Resonant regime with picoseconde laser excitation is realized by embedding the quantum dots in a waveguiding structure. As the pulse intensity is increased, Rabi oscillation is observed up to three periods. The Rabi regime is achieved owing to an enhanced light-matter coupling in the waveguide. This is due to a \emph{slow light effect} ($c/v_{g}\simeq 3000$), occuring when an intense resonant pulse propagates in a medium. The resonant control of the quantum dot fundamental transition opens new possibilities in quantum state manipulation and quantum optics experiments in condensed matter physics.
\end{abstract}

\maketitle

Manipulation of quantum bits, which are ideally two-level systems, is nowadays the subject to an important research effort knowing the possible applications in quantum computing. In condensed matter, the improvement of growth and integration techniques allows to think of elementary bricks for quantum computation based on solid state devices. Semiconductor quantum dots (QDs) appear to be very promising candidates since they have demonstrated similarities with simple atomic systems mainly because of their discrete energy spectrum. However interactions between the QD and its environment can cause a sever limitation of quantum coherence \cite{Wang}. Therefore in order to take full advantage of the QD states coherence, it is more interesting to manipulate the lowest lying transition since it is the least interacting with the environment and has the longest coherence time. An important feature of the strong non-linear interaction between a resonant field and a two level-system is Rabi oscillation (RO) in which the population of the excited state is driven by the field \cite{Cohen}. This is a key experiment for addressing and manipulating a given quantum state.\\
Furthermore, propagation experiments with ultrashort light pulses at resonance with a two-level system can result in coherent phenomena that modify the group velocity of light travelling in the medium, leading to \emph{pulse reshaping} \cite{Slusher} and \emph{self-induced transparency} \cite{McCall}, depending on different regimes of pulse area. Two physical processes produce slow light in atomic systems, \emph{coherent population oscillation} or \emph{electromagnetically induced transparency} \cite{Marangos}. The group velocity of light can be as slow as a few $m.s^{-1}$ \cite{Han} resulting in many investigations for storage of information or lasing without inversion of population. In semiconductor systems, similar coherent effects have been observed, although the physical processes are more complicated because of underlying many-body interactions. RO in quantum wells \cite{Cundiff}, pulse breakup and self-induced transparency \cite{Giessen} are some examples of these coherent non-linearities.

Here, we used a novel experimental configuration which allows to perform resonant excitation on the fundamental exciton state in a single QD and an all-optical control in real time of the driven state. Up to now, RO of an excited exciton state in a single QD has been performed \cite{Kamada}. Resonant RO of the fundamental transition in a single QD has recently been observed by means of four wave mixing \cite{Patton} or differential transmission \cite{Stievater} and indireclly by photocurrent measurements \cite{Stufler}. To our knowledge, resonant emission of a quantum state in a single QD  has never been reported because experimentally it is a real challenge to perform excitation and detection at the same wavelength. Our basic idea was to combine spatially resolved spectroscopy (microphotoluminescence, $\mu$PL) with a waveguiding geometry (Fig.1). The QDs of interest are embedded in an optical waveguide in which light propagates and interacts with the QDs. The emission of a spatially selected dot is collected in a perpendicular direction so the scattered laser intensity is low compared to the QD's luminescence. We demonstrate that in this geometry, excitation-induced nonlinearities of the excitonic resonance lead to an enhanced light-matter coupling allowing resonant RO of the excitonic state and simultaneously reducing the group velocity of the propagating light pulses.\\
The sample consists of QDs formed naturally by monolayer thickness fluctuations at the interfaces of a GaAs/GaAlAs quantum wire grown by metal organic vapor-phase epitaxy on a V-grooved GaAs substrate. These wires have been extensively studied \cite{Thierry1}. The QD's are elongated along the free wire axis with typical lengths from 10 to 100 nm. The number of dots along the wire is thus about 10 per micron as observed by scanning-imaging spectroscopy \cite{Thierry2}. The confinement potential along the free wire axis is about 10 meV leading to one or two discrete states inside the dots, depending on their size. \\
In order to form a \emph{monomode} waveguide, the GaAs quantum wire is embedded in AlGaAs layers with different Al concentrations to realize the core and the cladding of the waveguide. Due to the geometry of the structure, the optical modes have the same V-shape and the polarization of the optical modes has a particular distribution \cite{Dupertuis}. At the center of the V-shaped waveguide the two fundamental modes are along the Y and Z direction (Fig.1) and the polarization direction rotates toward the edges. The pump laser beam propagates along the free axis X of the wire and is usually polarized along the Y direction \cite{fils}.
\begin{figure}[b]
\centerline{\scalebox{0.6}{\includegraphics{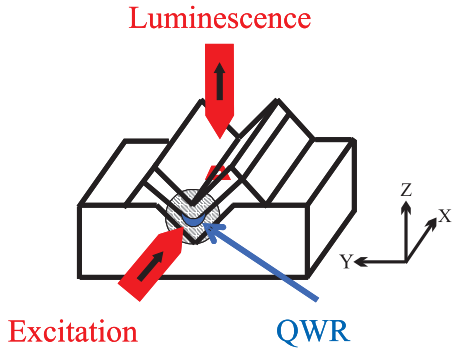}}} 
\caption{(color on-line) Schematic experimental configuration. The X-axis is the free axis of the wire and corresponds to the light propagation direction. The embedded QDs are elongated along the X axis. Y is the light polarization direction and Z is the growth axis. The grey circle and the blue crescent represent the optical mode and the quantum wire respectively.}\label{Guide}
\end{figure}
\\
A specific four optical accesses low drift cryostat has been designed especially for these experiments. The sample is fixed on a cold finger and cooled down to 10 K. The light source consists in a Ti-Sapphire laser with a few ps wide pulses. The laser beam can be focused through any of the high numerical aperture microscope objectives placed in front of the cryostat providing a diffraction limited spatial resolution of the order of $1 \mu m$. The beam is focused on the waveguide at the edge of the sample, then it propagates along the 1D waveguide pumping the different QDs embedded in it. The luminescence is collected in front of the sample by a microscope objective which can be moved along the waveguide axis, and a confocal geometry allows to probe locally one $\mu m^{2}$. The sample's luminescence is analyzed by a spectrometer with a spectral resolution of the order of $40 \mu eV$. The signal can then be detected either by a liquid nitrogen cooled CCD camera for spectroscopy or by a streak camera with a few ps temporal resolution for time-resolved measurements.
In order to achieve resonant excitation on a given excitonic transition, one of the crucials parameters is the structural quality of the waveguide interfaces. In fact, the inhomogeneities can induce leaks of the laser beam across the waveguiding structure and may completely blur the luminescence signal \cite{Cuba}.

The resonant emission of the fundamental exciton state in a QD (Fig.2) consists of a single lorentzian line with a FWHM of the order of 100 $\mu$eV, corresponding to a dephasing time of the order of 10 $ps$ for this particular QD. This homogeneous broadening can be explained by interaction with acoustic phonons that dephases the exciton state \cite{JAP}.
\begin{figure}[b]
\centerline{\scalebox{0.55}{\includegraphics{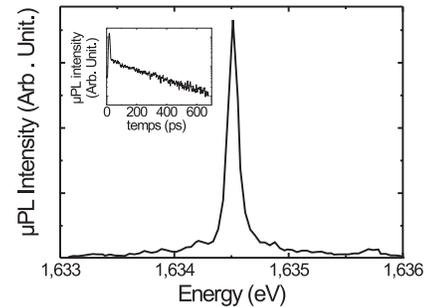}}} 
\caption{Resonant emission of a single QD at $10$K and low pump power. Inset: time-resolved luminescence excited at resonance. The decay time is 250 ps.}\label{SpcR}
\end{figure}
In the waveguiding geometry time-resolved luminescence of the fundamental exciton under resonant and non-resonant excitation has also been performed. Under non-resonant excitation the luminescence decay is mono-exponential and the exciton radiative lifetime is slightly longer than in the case of resonant excitation (inset of Fig.2). This shortening of the exciton decay can be understood by considering that the strong resonant excitation induces non-linear effects that act on the population of the excited state \cite{Chow}. Indeed, the waveguiding geometry enhances the absorption efficiency. We have also observed that the resonant exciton lifetime can vary with the pulse area. In the following we will discuss in more details these pecularities.\\
From the lifetime measurements we can estimate the QD length. In previous studies, we have shown that the radiative recombination rate is a linear function of the length of the dot \cite{Bellessa}. We find then for the studied QD, with a radiative lifetime of the order of 300 ps, a typical length of 40 nm. Andreani et al \cite{Andreani} calculated the oscillator strength (OS) for monolayer thickness fluctuations defects in GaAs/AlGaAs quantum wells with similar sizes to our QDs. By comparing the area of the dots we find that the OS associated to the fundamental excitonic transition is about $50$ and the dipole moment is also $50$ Debye. The obtained value is in agreement with other estimations in similar GaAs QDs and stronger than in the case of InAs self-assembled QDs \cite{Wang,Kamada,Guest}.

Under strong pulsed laser excitation, resonant with a two-level system, RO occurs as a function of the input pulse area (the time-integrated Rabi frequency) $\Theta=\int_{0}^{\infty}\Omega(t')dt'=\frac{\mu}{\hbar}\int_{0}^{\infty}E(t')dt'$, $E(t)$ being the pulse envelope, and $\mu$ the dipole moment of the transition \cite{Allen}. A complete oscillation between the two levels of the system is achieved when the pulse area is equal to $\Theta=2\pi$. Then the oscillations observed in the emission intensity scale like the $\sqrt{I}$ where $I$ is the intensity of the field related to the measured laser power as shown in Fig.3. Here we have plot the maximum of the resonant time-integrated $\mu$PL intensity as a function of the measured mean laser power.
\begin{figure}[b]
\centerline{\scalebox{0.6}{\includegraphics{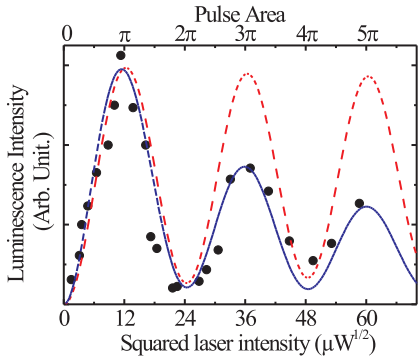}}} 
\caption{(color on-line) Rabi oscillations obtained from the integrated resonant luminescence (black dots) as a function of the pulse area. The blue solid line represents the calculated time-integrated population of the level and the red dashed curved represent the RO without damping parameter.}\label{SpcR}
\end{figure}
The spectral width of a few ps wide laser pulses is about 500 $\mu$eV covering all the single line emission of the exciton. Therefore, a broadened laser spectrum has been substracted in order to consider only the emission signal. Oscillations up to $5\pi$ have been observed in this particular QD corresponding to a pump power of a few mW. Beyond this value it becomes very difficult to distinguish between the emission and the scattered laser which becomes one of the main limitations in the manipulation of the state during the coherence regime. By using the semiclassical treatment of optical Bloch equations \cite{Mandel} we can fit the observed RO using a hyperbolic secant pulse. However, this simple two-level model does not predict the observed damping with increasing pulse area. Therefore we introduced a phenomenological intensity dependent damping parameter to account for the reduction of the oscillations amplitude. In this way, the time-integrated population of the excited state fits quite well the experimental data as shown in figure 3. The set of parameters used for the fit are: the population decay time $T_{1}$ (measured from the time-resolved luminescence) equal to 300 ps and the pulse duration of 1.5 ps. We have two ajustable parameters which are the coherence time $T_{2}$ (of the order of 25 ps) and the intensity dependant damping parameter. The value found for $T_{2}$ is a lower limit since to fit the data, the only requirement is that the pulse duration must be much smaller than $T_{2}$. \\
The presence of this phenomenological parameter is also consistent with the measured shortening of the exciton decay time as a function of pulse area. Indeed when we simulate the time evolution of the population we find the same order of variation of the decay time. The physical processes underlying this phenomenon can be non-linear effects like two-photon absorption or carrier-carrier scattering \cite{auger}. A Coulomb exciton-exciton scattering has also been proposed in \cite{Stievater}.

Surprisingly, when trying to fit the dipole moment of the transition from the relation $\Theta=\int_{0}^{\infty}\Omega(t')dt'=2\tau\frac{\mu E_{0}}{\hbar}$, with $E_{0}=\sqrt{P/\epsilon_{0}c}$  related to the measured laser power P (120 $\mu$W for a $\pi$-pulse) and with a pulse duration $\tau=$1.5 ps, we find a value higher by a factor 1000 than that estimated from the calculation discussed above. This is not consistent with what is usually expected for dipole moments in QDs. For the pump intensities used in our experiments, which are comparable to those used by other groups \cite{Kamada,Stievater,Stufler}, the only possibility to explain this result is to account for a much stronger light-matter coupling than the expected one.\\
It is well established that when an intense pulse travels in a medium at resonance, several non-linear effects can occur as a function of pulse area, acting both on the pulse and on the medium \cite{Boyd1}. In particular, at resonance, the refractive index varies very strongly due to a dispersive term $dn/d\omega$  leading to a modification of the group velocity of the propagating pulses. In the case of large positive dispersion, the group velocity becomes much smaller than the velocity of light in the medium ($v_{g}\ll c$). By Kramers-Kronig relations one finds that when the pulse intensity is increased, a spectral hole emerges in the absorption spectrum at resonance. All these effects are schematically depicted in Fig.4. 
\begin{figure}
\centerline{\scalebox{0.5}{\includegraphics{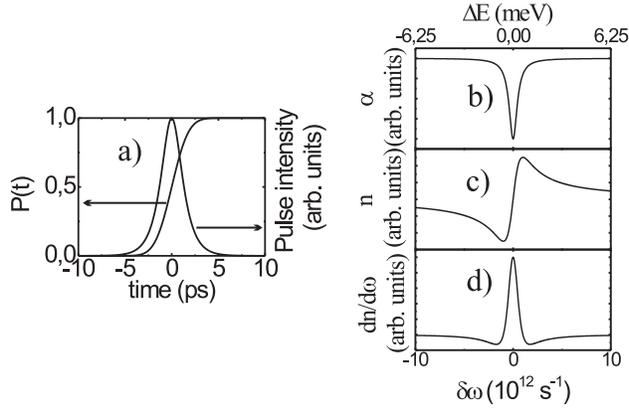}}} 
\caption{a) $\pi$-pulse shape and occupation probability of the excited state. Schematic representation of the frequency dependence of b) the absorption coefficient ($FT(1-P(t)$), c) the refractive index and d) the dispersion $dn/d\omega$.}\label{pulse}
\end{figure}
The transparency window around the resonance frequency is the origin for self-induced transparency observed for $2\pi$-pulses in atomic gases \cite{Slusher} or in semiconductor materials \cite{Giessen}. One can estimate the value of the group refractive index by assuming that the absorption coefficient of the medium is \cite{Boyd2} $\alpha(\omega)=\alpha_{0}\left(1-\frac{f}{1+(\omega-\omega_{0})^{2}/\gamma^{2}}\right)$, where $\alpha_{0}$ is the value of the background absorption, $\omega$ is the optical frequency, $\omega_{0}$ the resonance frequency and  the $\gamma$ the linewidth of the transparency window which is of the order of $1/\tau$. Indeed, qualitatively the change of the refractive index occurs during the interaction time with the pulse (Fig.4b). $f$ is a parameter that describes the transparency window, complete transparency is achieved for $f=1$. The refractive index associated to this absorption feature is \cite{Boyd2}: $n(\omega)=n_{0}+f\left(\frac{\alpha_{0}\lambda}{4\pi}\right)\frac{(\omega-\omega_{0})\tau}{1+(\omega-\omega_{0})^{2}\tau^{2}}$ and the group index is at resonance $n_{g}=n+\omega dn/d\omega\simeq\alpha_{0}\lambda\omega_{0}\tau/4\pi=\frac{\alpha_{0}L}{4\pi}\frac{\lambda}{L}\tau\omega_{0}$, where $\lambda$ is the vacuum wavelength and $L$ the interaction length equal, in our case, to the QD size. For a $\pi$-pulse, the absorption probability should be $\alpha_{0} L=1$, then the maxmum value of $n_{g}$ is 3000 for a QD length of 50 nm and the group velocity $v_{g}=c/n=10^{5} m.s^{-1}$. This corresponds to an absorption cross section $\sigma_{0}\simeq5.10^{-4}$ $\mu m^{-2}$ (for $\mu$=50 Debye), to compare to the 1 $\mu m^{2}$ section of the optical mode. Therefore for a $\pi$-pulse it is necessary to have $\sim$2000 photons to obtain an absorption probability equal to 1. The estimated number of photons agrees qualitatively well with the experimental excitation power for a $\pi$ pulse. 
This means that the pulse is delayed and experiences a spatial compression while interacting with the QD. Therefore the intensity of the coupling field is enhanced by the same factor.

In summary, resonant luminescence of a single QD has been directly observed. Light-matter interaction in the waveguiding geometry is clearly non-linear leading to a strong coupling regime allowing the achievement of RO. A compromise has to be found for the length of the QD in order to get both a high oscillator strength and a slow light effect strong enough to obtain an important light-matter coupling. In a similar way, the pulse duration has to be optimized so that the pulse spectral width fits that of the optical transition. The pulse delay after propagation in the medium is of the order of the pulse duration, therefore only for longer pulses would this effect be observable.

\begin{acknowledgments}
We thank M. Combescot and F. Dubin for stimulating discutions. This work has been supported by the Region Ile de France (SESAME No. E1751).
\end{acknowledgments}

\end{document}